%% file: main.tex
\pdfoutput=1

\documentclass[12pt,a4paper]{article}

\usepackage{ifthen} 
\newboolean{pdflatex}
\setboolean{pdflatex}{true} 

\newboolean{articletitles}
\setboolean{articletitles}{true} 

\newboolean{uprightparticles}
\setboolean{uprightparticles}{false} 

\input{macros}
\input{preamble}
\usepackage{booktabs}
\usepackage{rotating}

\begin{document}
\renewcommand{\thefootnote}{\fnsymbol{footnote}}
\setcounter{footnote}{1}
\input{title-LHCb-PAPER}

\renewcommand{\thefootnote}{\arabic{footnote}}
\setcounter{footnote}{0}


\pagestyle{plain} 
\setcounter{page}{1}
\pagenumbering{arabic}


\input{body}
\input{acknowledgements}

\addcontentsline{toc}{section}{References}
\bibliographystyle{LHCb}
\bibliography{main,standard,LHCb-PAPER,LHCb-CONF,LHCb-DP,LHCb-TDR}

\clearpage
\appendix
\include{supplemental-material}

 
\clearpage
\input{LHCb_AuthorsList/LHCb_Authorship_15-Jan-2019.tex}

\end{document}

%% file: macros.tex
\newcommand{\hhp}{\ensuremath{h^+h^{({\mkern-1mu\prime})-}}\xspace}
\newcommand{\PromptDecay}{\ensuremath{\Dstarp\to\Dz(\to\KS\pip\pim)\pip}\xspace}
\newcommand{\SLDecay}{\ensuremath{\Bbar\to\Dz(\to\KS\pip\pim)\mun X}\xspace}
\newcommand{\kspp}{\ensuremath{\KS\pip\pim}\xspace}
\newcommand{\Dkspp}{\ensuremath{\Dz\to\KS\pip\pim}\xspace}

\newcommand{\mD}{\ensuremath{m(\kspp)}\xspace}

\newcommand{\mpps}{\ensuremath{m^{2}(\pip\pim)}\xspace}

\newcommand{\zcp}{\ensuremath{z_{\CP}}\xspace}
\newcommand{\xcp}{\ensuremath{x_{\CP}}\xspace}
\newcommand{\ycp}{\ensuremath{y_{\CP}}\xspace}
\newcommand{\deltaz}{\ensuremath{\Delta z}\xspace}
\newcommand{\deltax}{\ensuremath{\Delta x}\xspace}
\newcommand{\deltay}{\ensuremath{\Delta y}\xspace}

\newcommand{\re}[2][()]{\ifthenelse{\equal{#1}{()}}{{\ensuremath{{\rm \, Re}}\!\left(#2\right)}}{{\ensuremath{{\rm \, Re}}\!\left[#2\right]}}}
\newcommand{\im}[2][()]{\ifthenelse{\equal{#1}{()}}{{\ensuremath{{\rm \, Im}}\!\left(#2\right)}}{{\ensuremath{{\rm \, Im}}\!\left[#2\right]}}}

\newcommand{\xcpRes}{\ensuremath{2.7}}
\newcommand{\xcpStat}{\ensuremath{1.6}}
\newcommand{\xcpSyst}{\ensuremath{0.4}}
\newcommand{\xUnits}{\times10^{-3}}
\newcommand{\ycpRes}{\ensuremath{7.4}}
\newcommand{\ycpStat}{\ensuremath{3.6}}
\newcommand{\ycpSyst}{\ensuremath{1.1}}
\newcommand{\yUnits}{\times10^{-3}}
\newcommand{\dxRes}{\ensuremath{-0.53}}
\newcommand{\dxStat}{\ensuremath{0.70}}
\newcommand{\dxSyst}{\ensuremath{0.22}}
\newcommand{\dxUnits}{\times10^{-3}}
\newcommand{\dyRes}{\ensuremath{0.6}}
\newcommand{\dyStat}{\ensuremath{1.6}}
\newcommand{\dySyst}{\ensuremath{0.3}}
\newcommand{\dyUnits}{\times10^{-3}}

\newcommand{\mypaperversion}{}
\newcommand{\mydate}{March 7, 2019}
\newcommand{\mylhcbpapernumber}{LHCb-PAPER-2019-001}
\newcommand{\mycernpapernumber}{CERN-EP-2019-032}

\def\paperauthors{LHCb collaboration} 
\def\paperasciititle{Measurement of the mass difference between neutral charm-meson eigenstates} 
\def\papertitle{Measurement of the mass difference between neutral charm-meson eigenstates} 
\def\paperkeywords{{High Energy Physics}, {LHCb}} 
\def\papercopyright{\the\year\ CERN for the benefit of the LHCb collaboration} 
\def\paperlicence{CC-BY-4.0 licence}
\def\paperlicenceurl{https://creativecommons.org/licenses/by/4.0/}

%% file: preamble.tex
\usepackage[top=1in, bottom=1.25in, left=1in, right=1in]{geometry}

\usepackage{microtype}
\usepackage{lineno}  
\usepackage{xspace} 
\usepackage{caption} 

\usepackage{graphicx}  
\usepackage{color}
\usepackage{colortbl}
\graphicspath{{./figs/}{./figs/paper/}{./figs/supplementary/}} 
\DeclareGraphicsExtensions{.pdf,.PDF,png,.PNG}

\usepackage{amsmath} 
\usepackage{amssymb}
\usepackage{amsfonts}
\usepackage{upgreek} 

\newcommand*\patchAmsMathEnvironmentForLineno[1]{%
\expandafter\let\csname old#1\expandafter\endcsname\csname #1\endcsname
\expandafter\let\csname oldend#1\expandafter\endcsname\csname
end#1\endcsname
 \renewenvironment{#1}%
   {\linenomath\csname old#1\endcsname}%
   {\csname oldend#1\endcsname\endlinenomath}%
}
\newcommand*\patchBothAmsMathEnvironmentsForLineno[1]{%
  \patchAmsMathEnvironmentForLineno{#1}%
  \patchAmsMathEnvironmentForLineno{#1*}%
}
\AtBeginDocument{%
\patchBothAmsMathEnvironmentsForLineno{equation}%
\patchBothAmsMathEnvironmentsForLineno{align}%
\patchBothAmsMathEnvironmentsForLineno{flalign}%
\patchBothAmsMathEnvironmentsForLineno{alignat}%
\patchBothAmsMathEnvironmentsForLineno{gather}%
\patchBothAmsMathEnvironmentsForLineno{multline}%
\patchBothAmsMathEnvironmentsForLineno{eqnarray}%
}

\usepackage{hyperxmp}

\usepackage[pdftex,
            pdfauthor={\paperauthors},
            pdftitle={\paperasciititle},
            pdfkeywords={\paperkeywords},
            pdfcopyright={Copyright (C) \papercopyright},
            pdflicenseurl={\paperlicenceurl}]{hyperref}

\usepackage[colorinlistoftodos,textsize=scriptsize]{todonotes}

\usepackage[all]{hypcap} 

\input{lhcb-symbols-def} 

\usepackage{cite} 
\usepackage{mciteplus}

\usepackage[capitalise]{cleveref}
\Crefname{figure}{Figure}{Figures}

%% file: lhcb-symbols-def.tex
\usepackage{xspace} 
\usepackage{upgreek}

\newcommand{\offsetoverline}[2][0.1em]{\kern #1\overline{\kern -#1 #2}}%

\def\lhcb   {\mbox{LHCb}\xspace}

\def\MagUp {\mbox{\em Mag\kern -0.05em Up}\xspace}

\ifthenelse{\boolean{uprightparticles}}%
{

 \def\Pmu         {\ensuremath{\upmu}\xspace}

 \def\Ppi         {\ensuremath{\uppi}\xspace}

 \def\PDelta      {\ensuremath{\Delta}\xspace}                 
 \def\PXi         {\ensuremath{\Xi}\xspace}                 
 \def\PLambda     {\ensuremath{\Lambda}\xspace}                 
 \def\PSigma      {\ensuremath{\Sigma}\xspace}                 
 \def\POmega      {\ensuremath{\Omega}\xspace}                 
 \def\PUpsilon    {\ensuremath{\Upsilon}\xspace}

 \def\PB      {\ensuremath{\mathrm{B}}\xspace}                 
                  
 \def\PD      {\ensuremath{\mathrm{D}}\xspace}

 \def\PK      {\ensuremath{\mathrm{K}}\xspace}

 \def\Pb      {\ensuremath{\mathrm{b}}\xspace}                 
 \def\Pc      {\ensuremath{\mathrm{c}}\xspace}

 \def\Pi      {\ensuremath{\mathrm{i}}\xspace}

}
{

 \def\Pmu         {\ensuremath{\mu}\xspace}

 \def\Ppi         {\ensuremath{\pi}\xspace}

 \mathchardef\PDelta="7101
 \mathchardef\PXi="7104
 \mathchardef\PLambda="7103
 \mathchardef\PSigma="7106
 \mathchardef\POmega="710A
 \mathchardef\PUpsilon="7107
                  
 \def\PB      {\ensuremath{B}\xspace}                 
                  
 \def\PD      {\ensuremath{D}\xspace}

 \def\PK      {\ensuremath{K}\xspace}

 \def\Pb      {\ensuremath{b}\xspace}                 
 \def\Pc      {\ensuremath{c}\xspace}

 \def\Pi      {\ensuremath{i}\xspace}

}

\makeatletter
\ifcase \@ptsize \relax
  \newcommand{\miniscule}{\@setfontsize\miniscule{4}{5}}
\or
  \newcommand{\miniscule}{\@setfontsize\miniscule{5}{6}}
\or
  \newcommand{\miniscule}{\@setfontsize\miniscule{5}{6}}
\fi
\makeatother

\DeclareRobustCommand{\optbar}[1]{\shortstack{{\miniscule (\rule[.5ex]{1.25em}{.18mm})}
  \\ [-.7ex] $#1$}}

\def\mun        {{\ensuremath{\Pmu^-}}\xspace} 

\def\cquark    {{\ensuremath{\Pc}}\xspace}

\def\bquark    {{\ensuremath{\Pb}}\xspace}

\def\pion   {{\ensuremath{\Ppi}}\xspace}

\def\pip    {{\ensuremath{\pion^+}}\xspace}
\def\pim    {{\ensuremath{\pion^-}}\xspace}

\def\kaon    {{\ensuremath{\PK}}\xspace}
  \def\Kbar    {{\kern 0.2em\overline{\kern -0.2em \PK}{}}\xspace}

\def\KorKbar {\kern 0.18em\optbar{\kern -0.18em K}{}\xspace}

\def\KS      {{\ensuremath{\kaon^0_{\mathrm{S}}}}\xspace}

\def\Kstar   {{\ensuremath{\kaon^*}}\xspace}

  \def\Dbar    {{\kern 0.2em\overline{\kern -0.2em \PD}{}}\xspace}
\def\D       {{\ensuremath{\PD}}\xspace}

\def\DorDbar {\kern 0.18em\optbar{\kern -0.18em D}{}\xspace}
\def\Dz      {{\ensuremath{\D^0}}\xspace}
\def\Dzb     {{\ensuremath{\Dbar{}^0}}\xspace}

\def\Dstar   {{\ensuremath{\D^*}}\xspace}

\def\Dstarp  {{\ensuremath{\D^{*+}}}\xspace}

\def\Bbar    {{\ensuremath{\kern 0.18em\overline{\kern -0.18em \PB}{}}}\xspace}
\def\Bb      {{\ensuremath{\Bbar}}\xspace}
\def\BorBbar    {\kern 0.18em\optbar{\kern -0.18em B}{}\xspace}

\def\Y#1S{\ensuremath{\PUpsilon{(#1S)}}\xspace}

\def\LorLbar     {\kern 0.18em\optbar{\kern -0.18em \PLambda}{}\xspace}

\newcommand{\decay}[2]{\mbox{\ensuremath{#1\!\to #2}}\xspace}         

\def\to                 {\ensuremath{\rightarrow}\xspace}

\def\CP                {{\ensuremath{C\!P}}\xspace}

\newcommand{\dm}{{\ensuremath{\Delta m}}\xspace}

\def\AT#1     {\ensuremath{A_{\mathrm{T}}^{#1}}\xspace}           

\def\C#1      {\ensuremath{\mathcal{C}_{#1}}\xspace}                       
\def\Cp#1     {\ensuremath{\mathcal{C}_{#1}^{'}}\xspace}                    
\def\Ceff#1   {\ensuremath{\mathcal{C}_{#1}^{\mathrm{(eff)}}}\xspace}        
\def\Cpeff#1  {\ensuremath{\mathcal{C}_{#1}^{'\mathrm{(eff)}}}\xspace}       
\def\Ope#1    {\ensuremath{\mathcal{O}_{#1}}\xspace}                       
\def\Opep#1   {\ensuremath{\mathcal{O}_{#1}^{'}}\xspace}                    

\def\ycp        {\ensuremath{y_{\CP}}\xspace}

\newcommand{\aunit}[1]{\ensuremath{\text{\,#1}}}       

\newcommand{\tev}{\aunit{Te\kern -0.1em V}\xspace}
\newcommand{\gev}{\aunit{Ge\kern -0.1em V}\xspace}
\newcommand{\mev}{\aunit{Me\kern -0.1em V}\xspace}
\newcommand{\kev}{\aunit{ke\kern -0.1em V}\xspace}
\newcommand{\ev}{\aunit{e\kern -0.1em V}\xspace}
\newcommand{\mevc}{\ensuremath{\aunit{Me\kern -0.1em V\!/}c}\xspace}
\newcommand{\gevc}{\ensuremath{\aunit{Ge\kern -0.1em V\!/}c}\xspace}
\newcommand{\mevcc}{\ensuremath{\aunit{Me\kern -0.1em V\!/}c^2}\xspace}
\newcommand{\gevcc}{\ensuremath{\aunit{Ge\kern -0.1em V\!/}c^2}\xspace}

\def\fb   {\ensuremath{\aunit{fb}}\xspace}
\def\invfb   {\ensuremath{\fb^{-1}}\xspace}

\newcommand{\stat}{\aunit{(stat)}\xspace}
\newcommand{\syst}{\aunit{(syst)}\xspace}

\newcommand{\chisq}{\ensuremath{\chi^2}\xspace}

\def\gsim{{~\raise.15em\hbox{$>$}\kern-.85em
          \lower.35em\hbox{$\sim$}~}\xspace}
\def\lsim{{~\raise.15em\hbox{$<$}\kern-.85em
          \lower.35em\hbox{$\sim$}~}\xspace}

\xspace

\def\tell1  {TELL1\xspace}
\def\ukl1   {UKL1\xspace}



%% file: title-LHCb-PAPER.tex

\begin{titlepage}
\pagenumbering{roman}

\vspace*{-1.5cm}
\centerline{\large EUROPEAN ORGANIZATION FOR NUCLEAR RESEARCH (CERN)}
\vspace*{1.5cm}
\noindent
\begin{tabular*}{\linewidth}{lc@{\extracolsep{\fill}}r@{\extracolsep{0pt}}}
\vspace*{-1.5cm}\mbox{\!\!\!\includegraphics[width=.14\textwidth]{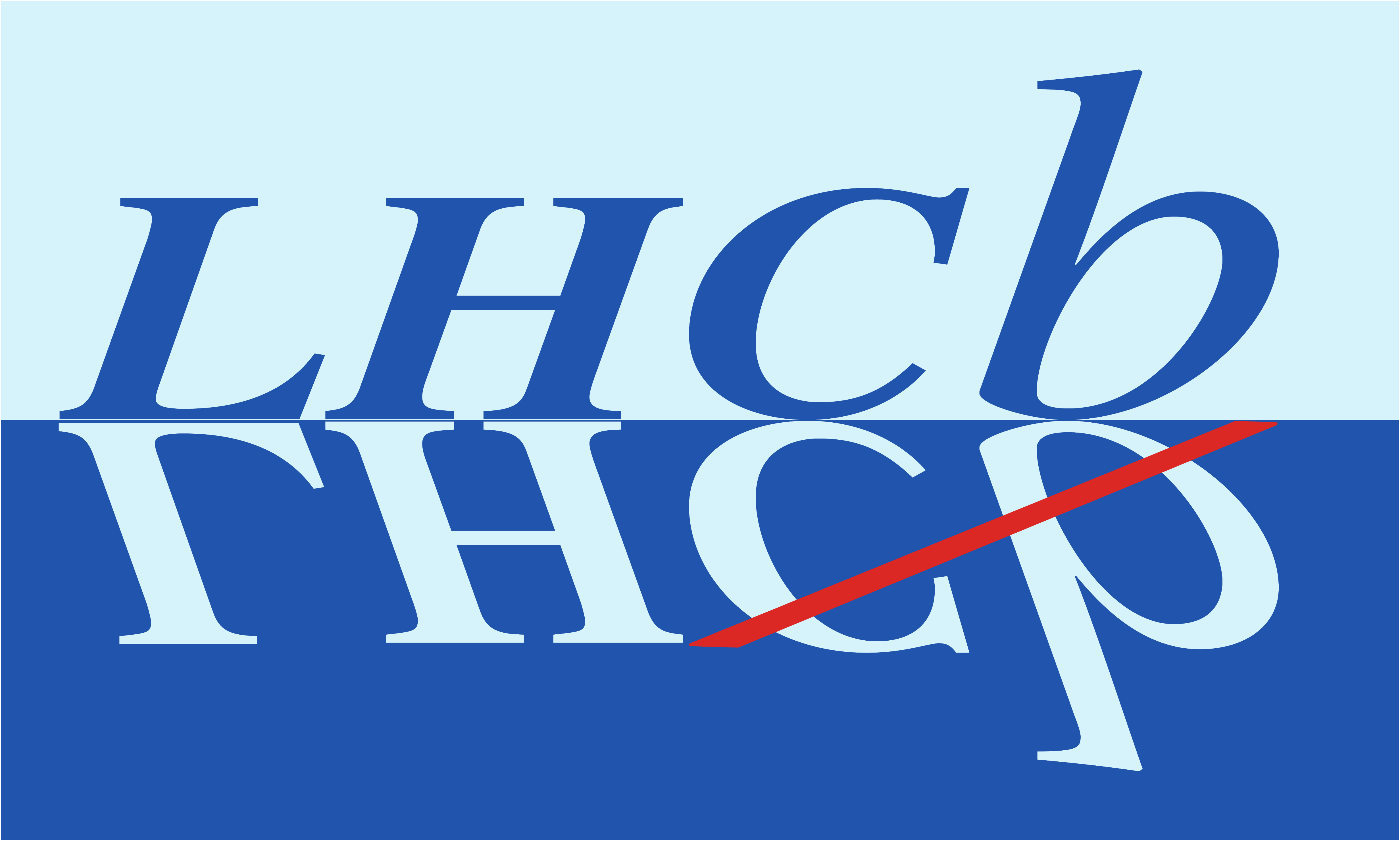}} & & \\
 & & \mycernpapernumber \\
 & & \mylhcbpapernumber \\
 & & \mydate \\ 
 & & \mypaperversion\\
\end{tabular*}

\vspace*{4.0cm}

{\normalfont\bfseries\boldmath\huge
\begin{center}
  \papertitle 
\end{center}
}

\vspace*{1.5cm}

\begin{center}
\paperauthors\footnote{Authors are listed at the end of this paper.}
\end{center}

\vspace{\fill}

\begin{abstract}
\input{abstract}
\end{abstract}

\vspace*{1.5cm}

\begin{center}
Published in Phys.\ Rev.\ Lett.\ {\bf 122} (2019) 231802
\end{center}

\vspace{\fill}

{\footnotesize 
\centerline{\copyright~\papercopyright. \href{\paperlicenceurl}{\paperlicence}.}}
\vspace*{2mm}

\end{titlepage}


\newpage
\setcounter{page}{2}
\mbox{~}

\cleardoublepage

%% file: abstract.tex
\noindent
We report a measurement of the mass difference between neutral charm-meson eigenstates using a novel approach that enhances sensitivity to this parameter. We use $2.3\times 10^6$ \Dkspp decays reconstructed in proton-proton collisions collected by the \lhcb experiment in 2011 and 2012. Allowing for \CP violation in mixing and in the interference between mixing and decay, we measure the \CP-averaged normalized mass difference $\xcp = \left[\xcpRes\pm\xcpStat\stat\pm\xcpSyst\syst\right]\xUnits$ and the \CP-violating parameter $\deltax = \left[\dxRes\pm\dxStat\stat\pm\dxSyst\syst\right]\dxUnits$. The results are consistent with \CP symmetry. These determinations are the most precise from a single experiment and, combined with current world-average results, yield the first evidence that the masses of the neutral charm-meson eigenstates differ.

%% file: body.tex
Flavor oscillations are transitions between neutral flavored mesons and their corresponding antimesons that follow an oscillating pattern as a function of decay time. In the standard model, these transitions are mediated by weak-interaction amplitudes involving exchanges of virtual $W^\pm$ bosons and heavy quarks. Unknown particles of arbitrarily high mass can contribute as virtual particles in the amplitude, possibly enhancing the average oscillation rate or the difference between the rates of mesons and antimesons. This makes flavor oscillations sensitive to non-standard-model dynamics at large energy scales~\cite{Isidori:2010kg}.

Oscillations occur because the mass eigenstates of neutral flavored mesons are linear combinations of the flavor eigenstates. In particular, for charm mesons one writes \mbox{$|D_{1,2}\rangle\equiv p|\Dz\rangle\pm q |\Dzb\rangle$}, where $p$ and $q$ are complex parameters. In the limit of charge-parity (\CP) symmetry, and by defining $D_{1(2)}$ as the \CP-even (odd) state, the oscillation rate depends only on the dimensionless mixing parameters \mbox{$x\equiv(m_1 - m_2)c^2/\Gamma$} and \mbox{$y\equiv(\Gamma_1 - \Gamma_2)/(2\Gamma)$}, where $m_{1(2)}$ and $\Gamma_{1(2)}$ are the mass and decay width of the $D_{1(2)}$ state, respectively, and $\Gamma$ equals $(\Gamma_1+\Gamma_2)/2$~\cite{PDG2018}. If \CP symmetry is violated, the oscillation rates for mesons produced as \Dz and \Dzb differ. The difference is generated in the mixing amplitude if $|q/p|\neq1$ or in the interference between mixing and decay if \mbox{$\phi_f\equiv\arg(q\bar{A}_f/pA_f)\neq0$}. The amplitude $A_f$ ($\bar{A}_f$) refers to the decay $\Dz\to f$ ($\Dzb\to f$), where $f$ is a common final state. If \CP is conserved in  the decay amplitude ($|A_f|^2=|\bar{A}_f|^2$), the \CP-violating phase is independent of the final state, $\phi_f\approx\phi=\arg(q/p)$~\cite{Du:1986ai,Bergmann:2000id}.

Current global averages of charm-mixing parameters have large uncertainties and are consistent with \CP symmetry, yielding $x = (3.6\,^{+\,2.1}_{-\,1.6})\times10^{-3}$, $y = (6.7\,^{+\,0.6}_{-\,1.3})\times10^{-3}$, $|q/p|=0.94\,^{+\,0.17}_{-\,0.07}$, and $\phi=-0.13\,^{+\,0.26}_{-\,0.17}$~\cite{HFLAV16}. Improving the knowledge of $x$, which has not been shown to differ significantly from zero, is especially critical because the sensitivity to the small phase $\phi$ relies predominantly on observables proportional to $x\sin\phi$.

Direct experimental access to charm-mixing parameters is offered by self-conjugate multibody decays, such as \Dkspp. Inclusion of charge-conjugate processes is implied unless stated otherwise. A joint fit of the Dalitz-plot and decay-time distributions of these decays allows the identification of a \Dz component that increases as a function of decay time in a sample of candidates produced as \Dzb mesons, and vice versa. This approach is challenging because it requires analyzing the decay-time evolution of signal decays across the Dalitz plot with a detailed amplitude model while accounting for efficiencies, resolutions, and background~\cite{Asner:2005sz,Peng:2014oda,delAmoSanchez:2010xz}. Model-independent approaches that obviate the need for an amplitude analysis exist~\cite{Bondar:2010qs,Thomas:2012qf,LHCb-PAPER-2015-042}, but they rely on an accurate description of the efficiencies.

This Letter reports on a measurement of charm oscillations in \mbox{\Dkspp} decays based on a novel model-independent approach, called the bin-flip method, which is optimized for the measurement of the parameter $x$~\cite{binflip-paper}. The method relies on ratios between charm decays reconstructed in similar kinematic and decay-time conditions, thus avoiding the need for an accurate modeling of the efficiency variation across phase space and decay time. We express the \Dkspp dynamics with two invariant masses following the Dalitz formalism~\cite{Dalitz:1953cp,Fabri:1954zz}, where $m_\pm^2$ is the squared invariant mass $m^2(\KS\pi^\pm)$ for $\Dz\to\kspp$ decays and $m^2(\KS\pi^\mp)$ for $\Dzb\to\kspp$ decays. We partition the Dalitz plot into disjoint regions (``bins'') that preserve nearly constant strong-phase differences $\Delta\delta(m_-^2,m_+^2)$ between the \Dz and \Dzb amplitudes within each bin (see \nameref{supplemental-material}). Two sets of eight bins are formed, and they are organized symmetrically about the principal bisector $m_+^2 = m_-^2$. Bins are labeled with the indices $\pm b$, where $b=1,...,8$. Positive indices refer to the (lower) $m_+^2 > m_-^2$ region, where unmixed Cabibbo-favored \decay{\Dz}{\Kstar(892)^-\pi^+} decays dominate; negative indices refer to the symmetric (upper) $m_+^2 < m_-^2$ region, which receives a larger contribution from decays following oscillation. The data are further split into bins of decay time, which are indexed with $j$. For each, we measure the ratio $R_{bj}^+$ ($R_{bj}^-$) between initially produced \Dz (\Dzb) mesons in Dalitz bin $-b$ and Dalitz bin $b$. For small mixing parameters and \CP-conserving decay amplitudes, which are good approximations here, the ratios are~\cite{binflip-paper}
\begin{equation}\label{eq:bin-flip-ratio}
R_{bj}^\pm \approx \frac{r_b + \dfrac{1}{4}\,r_b\,\langle t^2\rangle_j\re{\zcp^2-\deltaz^2} + \dfrac{1}{4}\,\langle t^2\rangle_j\left|\zcp\pm\deltaz\right|^2 + \sqrt{r_b}\langle t\rangle_j \re[]{X_b^*(\zcp\pm\deltaz)}}{1 + \dfrac{1}{4}\,\langle t^2\rangle_j\re{\zcp^2-\deltaz^2}+r_b\,\dfrac{1}{4}\,\langle t^2\rangle_j\left|\zcp\pm\deltaz\right|^2 + \sqrt{r_b} \langle t\rangle_j \re[]{X_b(\zcp\pm\deltaz)}}.
\end{equation}
Here $\langle t\rangle_j$ ($\langle t^2\rangle_j$) is the average (squared) decay time of unmixed decays in bin $j$, in units of the \Dz lifetime $\tau=\hbar/\Gamma$~\cite{PDG2018}. The parameter $r_b$ is the ratio of signal yields in symmetric Dalitz-plot bins $\mp b$ at $t=0$, and $X_b$ quantifies the average strong-phase difference in these bins~\cite{binflip-paper}. The \zcp and \deltaz parameters, defined by $\zcp\pm\deltaz\equiv-\left(q/p\right)^{\pm1}(y+ix)$, are obtained, along with $r_b$, from a joint fit of the observed $R_{bj}^\pm$ ratios in which external information on $c_b\equiv\re{X_b}$ and $s_b\equiv-\im{X_b}$~\cite{Libby:2010nu} is used as a constraint. The results are expressed in terms of the \CP-averaged mixing parameters \mbox{$\xcp\equiv-\im{\zcp}$} and \mbox{$\ycp\equiv-\re{\zcp}$}, and of the \CP-violating differences \mbox{$\deltax\equiv-\im{\deltaz}$} and \mbox{$\deltay\equiv-\re{\deltaz}$}. Conservation of \CP symmetry in mixing, or in the interference between mixing and decay, implies $\xcp=x$, $\ycp=y$, and $\deltax=\deltay=0$.

Samples of \Dkspp decays are reconstructed from proton-proton collisions collected by the \lhcb experiment in 2011 and 2012, corresponding to integrated luminosities of 1\invfb and 2\invfb, respectively. In the 2012 data, both the strong-interaction decay $\Dstarp\to\Dz\pip$ and the semileptonic \bquark-hadron decay $\Bbar\to\Dz\mun X$, where $X$ generically indicates unreconstructed particles, are used to determine whether a \Dz or a \Dzb is produced. In the 2011 data, only the $\Bbar\to\Dz\mun X$ decays were used because the online-selection efficiency for $\Dstarp\to\Dz\pip$ decays was low. Throughout this Letter, \Dstarp indicates the $\Dstar(2010)^+$ meson and soft pion indicates the pion from its decay.

The \lhcb detector is a single-arm forward spectrometer covering the \mbox{pseudorapidity} range $2<\eta<5$ equipped with charged-hadron identification detectors, calorimeters, and muon detectors; and it is designed for the study of particles containing \bquark or \cquark quarks~\cite{Alves:2008zz,LHCb-DP-2014-002}.

The online selection of \PromptDecay decays (prompt sample) uses criteria on momenta and final-state charged-particle displacements from any proton-proton primary interaction. Offline, we apply criteria consistent with the decay topology on momenta, vertex and track displacements, particle-identification information, and invariant masses of the \Dstarp decay products. Specifically, the mass of the \Dz candidate is required to meet $1.84<\mD<1.89\gevcc$ and the difference between the \Dstarp and \Dz candidate masses is required to satisfy $\dm<151.1\mevcc$. The \Dz and soft pion candidates are required to point back to one of the proton-proton interactions (the primary vertex) to suppress signal candidates originating from decays of \bquark hadrons (secondary decays). A kinematic fit constrains the tracks according to the decay topology and the \Dstarp candidate to originate from the primary vertex~\cite{Hulsbergen:2005pu}. In the reconstruction of the Dalitz-plot coordinates, we additionally constrain the \KS and \Dz meson masses to the known values~\cite{PDG2018} to ensure that all candidates populate the kinematically allowed phase space.

The online selection of \SLDecay decays (semileptonic sample) requires at least one displaced high-transverse-momentum muon and a vertex consistent with the decay of a \bquark hadron. Offline, we apply criteria consistent with the decay topology on momenta, vertex and track displacements, particle identifications, and invariant masses of the \Dz decay products. In addition, candidate $\Dz\mun$ pairs are formed by requiring \mbox{$2.5 < m(\Dz\mun)<6.0\gevcc$} and the corrected mass \mbox{$\sqrt{m^2(\Dz\mun)+p_\perp^2(\Dz\mun)}+p_\perp(\Dz\mun)$}, where the momentum component $p_\perp(\Dz\mun)$ of the $\Dz\mun$ system transverse to the \Bb flight direction partially compensates for the momentum of unreconstructed decay products, to be smaller than $5.8\gevcc$. The \Bb flight direction is inferred from the measured positions of the primary and $\Dz\mun$ vertices. A kinematic fit constrains the \Dz and \KS masses to their known values.

In both samples, two categories of signal candidates are used, those with \decay{\KS}{\pip\pim} candidates reconstructed in the vertex detector (long \KS) and those with \KS candidates reconstructed after the vertex detector (downstream \KS).

About 2\% (3\%) of the selected \Dstarp (\Bb) candidates belong to events in which multiple candidates are reconstructed by pairing the same \Dz candidate with different soft pions (muons). For these events, we randomly choose a single candidate. We consider the prompt and semileptonic samples independent because their overlap amounts to less than 0.1\% of the semileptonic sample size.

\begin{figure}[t]
\centering
\includegraphics[width=0.5\textwidth]{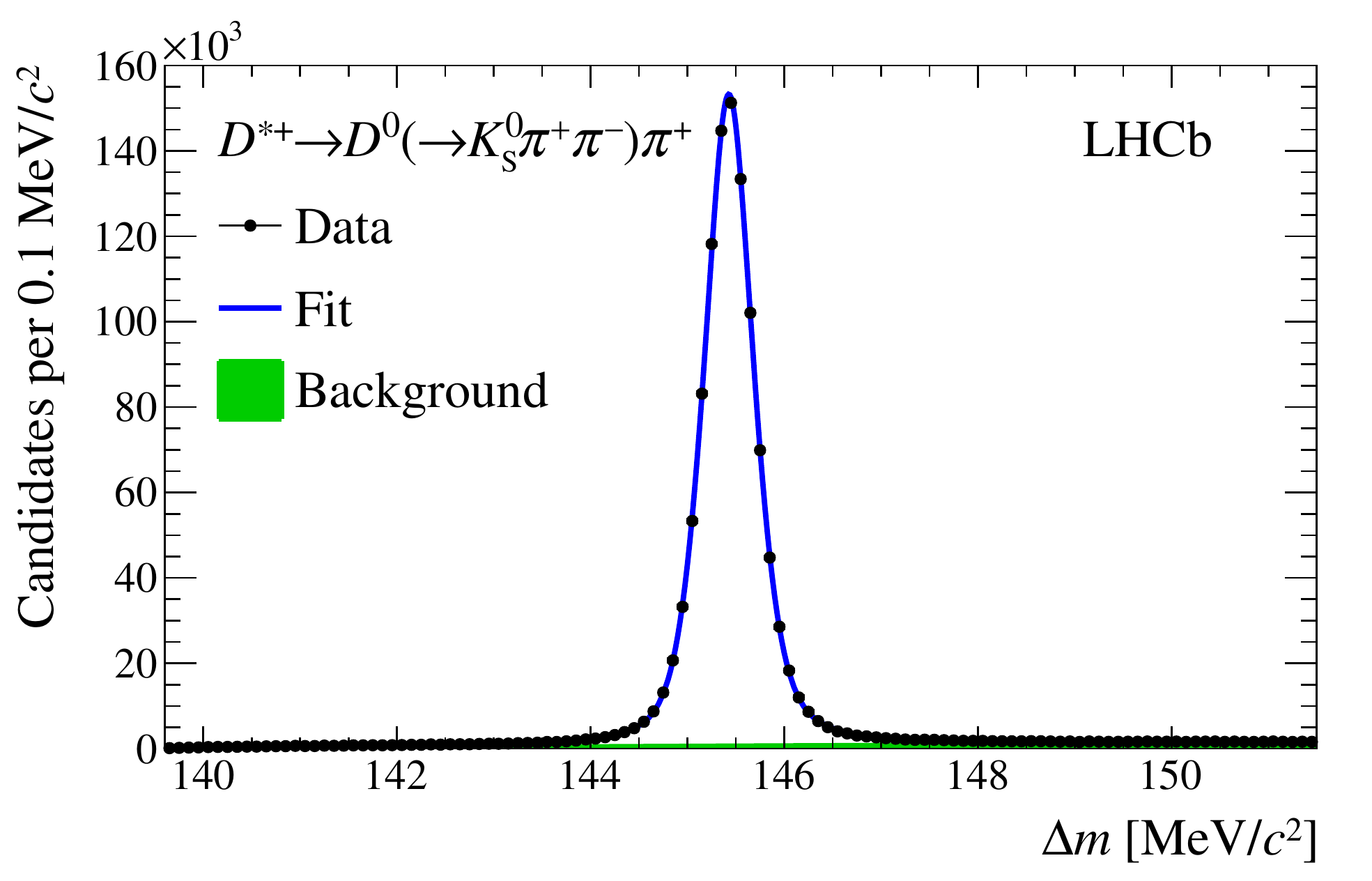}\hfil
\includegraphics[width=0.5\textwidth]{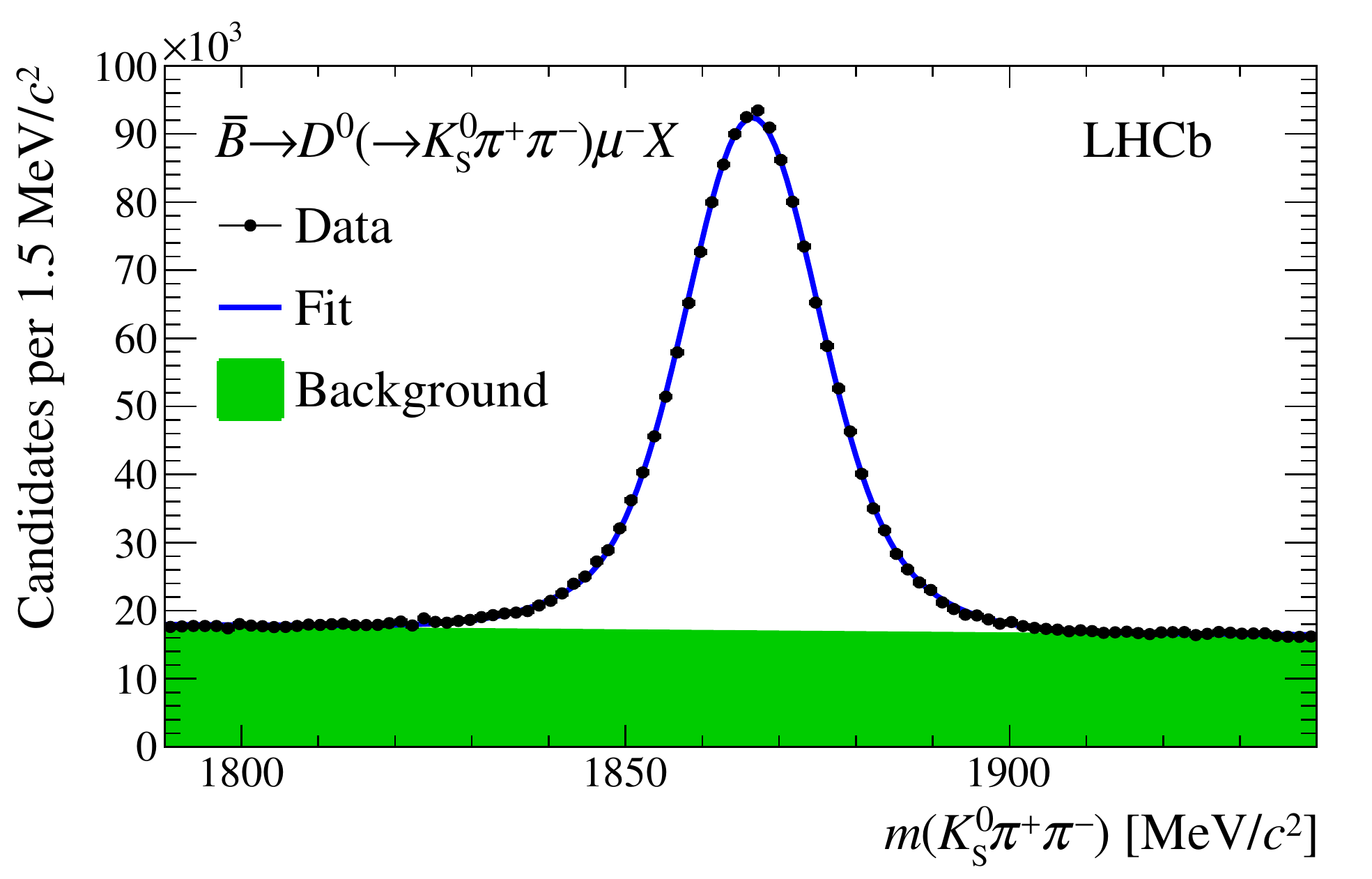}\\
\caption{Distribution of (left) the difference between \Dstarp and \Dz masses for \mbox{\PromptDecay} candidates and (right) \Dz mass for \SLDecay candidates.
\label{fig:example-fits}}
\end{figure}

\Cref{fig:example-fits} shows the \dm and \mD distributions of the prompt and semileptonic samples, respectively. The prompt sample contains $1.3 \times 10^6$ signal decays (45\% with downstream \KS candidates) and a small background dominated by genuine \Dkspp decays associated to random soft pions. Secondary \Dstarp decays contribute approximately 3\% to the signal yield, as determined using  \Dz candidates not pointing to the primary vertex. The semileptonic sample contains $1.0\times10^6$ signal decays ($66\%$ with downstream \KS candidates) and a sizable background dominated by unrelated \kspp combinations. Genuine \Dz decays associated with random muons contribute less than 1\% to the \Dz yield, as determined from the yield of false \Bbar candidates formed by associating $\Dstarp\to\Dz\pip$ with same-sign $\mu^+$ candidates. Contributions from backgrounds due to misreconstructed \Dz decays, such as $\Dz\to\KS\pip\pim\pi^0$ and $\Dz\to\KS\hhp$, where \hhp indicates a pair of light hadrons other than $\pi^+\pi^-$, are negligible. 

\begin{figure}[t]
\centering
\includegraphics[width=0.6\textwidth]{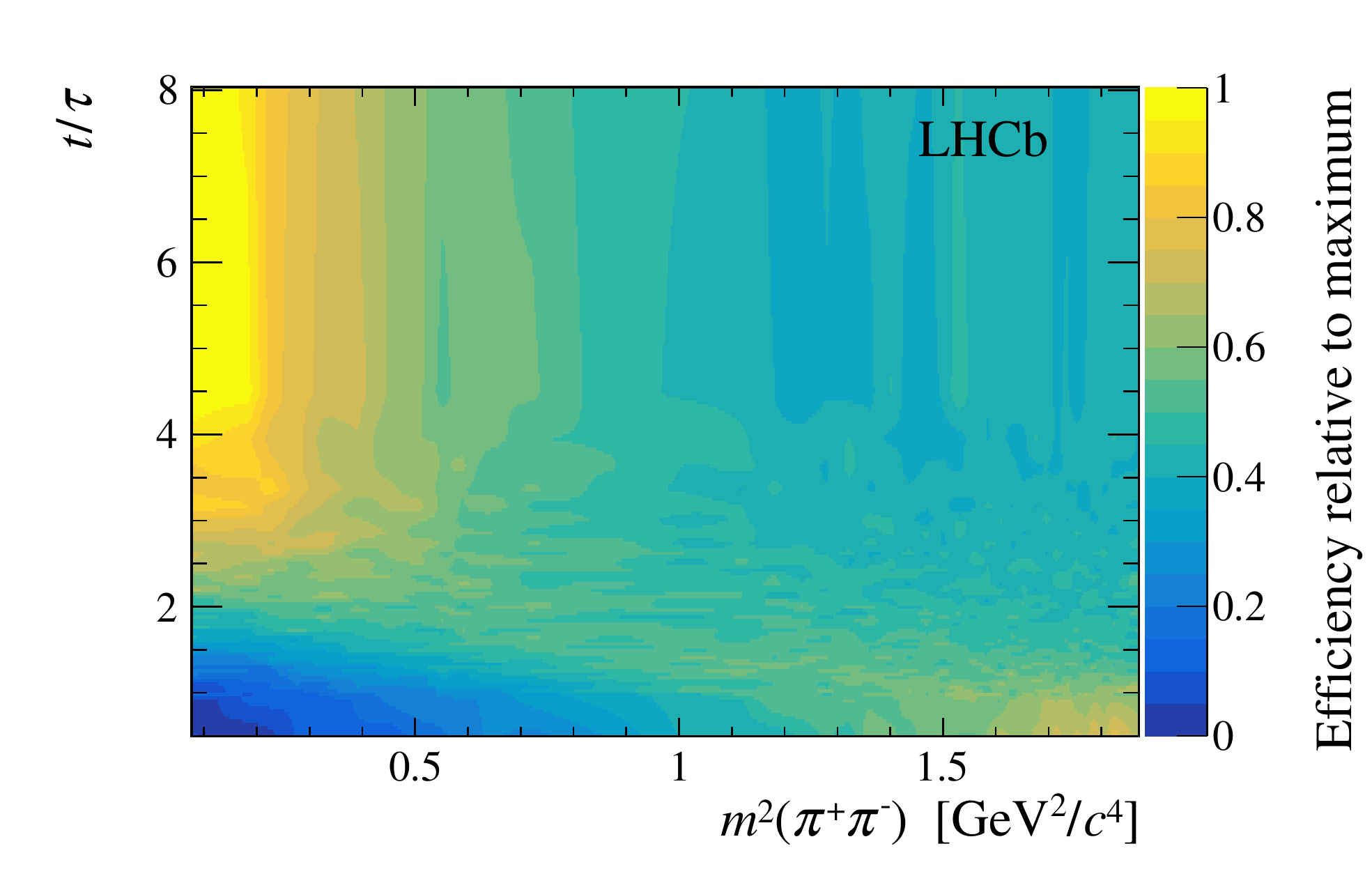}\\
\caption{Smoothed efficiency as a function of \mpps and $t/\tau$ in \PromptDecay decays, as determined from the data with downstream \KS candidates.\label{fig:eff_corr_prompt_data}}
\end{figure}

Simulated~\cite{LHCb-PROC-2011-006,LHCb-PROC-2010-056} prompt decays show that the online requirements on displacement and momenta of the \Dz decay products introduce efficiency variations that are correlated between the squared mass of the two final-state pions, \mpps, and the \Dz decay time. Because $\left(\mpps,t\right)$ correlations can bias the results, we correct for them using data. The smallness of the mixing parameters~\cite{HFLAV16}, along with the known \Dkspp decay amplitudes~\cite{Asner:2005sz,Peng:2014oda,delAmoSanchez:2010xz}, rules out any measurable $\left(\mpps,t\right)$ correlation introduced by \Dz--\Dzb mixing with current sample sizes. Hence, we ascribe any observed dependence between \mpps and $t$ to instrumental effects. We use the background-subtracted $\left(\mpps,t\right)$ distribution to determine the decay-time efficiency, normalized  to the average decay-time distribution, as a function of \mpps. This two-dimensional map is smoothed and used to assign per-candidate weights proportional to the inverse of the relative efficiency at each candidate's $\left(\mpps,t\right)$ coordinates, effectively removing the correlated nonuniformities. The corrections are determined separately for long and downstream \KS candidates because they feature different correlations. \Cref{fig:eff_corr_prompt_data} shows the smoothed $\left(\mpps,t\right)$ map for the sample with downstream \KS candidates, where the correlations are more prominent. The 6\% of candidates reconstructed with $t<0.9\tau$ are discarded because the corresponding weights cannot be determined precisely. No $\left(\mpps,t\right)$ correlations are observed in \SLDecay decays.

We divide prompt and semileptonic samples according to the \KS category, \Dz meson flavor, Dalitz-plot position, and decay time. In each subsample, we determine the signal yield and --- for each decay-time bin --- the average decay time and average squared decay time of the signal candidates. Finally, we fit the decay-time dependence of the ratio of the signal yields symmetric with respect to the Dalitz-plot bisector.

We determine the signal yields by fitting the \dm distribution, weighted to correct for the $\left(\mpps,t\right)$ correlations, for the \PromptDecay candidates and the \mD distribution for the \SLDecay candidates. All components are modeled empirically. The \dm model combines a \Dstarp signal with a smooth phase-space-like background. The \mD model combines a \Dz signal with a linear background. Signal and background shape parameters are determined independently for long and downstream \KS candidates, for \Dz and \Dzb mesons, and in each decay-time and Dalitz-plot bin. The signal model assumes the same parameters for each pair of positive and negative Dalitz-plot bins.

We estimate $\langle t \rangle_j$ and $\langle t^2 \rangle_j$ from the background-subtracted $t$ distribution in each decay-time bin $j$ separately for prompt and semileptonic samples, as well as for long and downstream \KS candidates. Background is subtracted using weights derived from the mass fits~\cite{Pivk:2004ty} of candidates restricted to the lower half ($m^2_{-}<m^2_{+}$) of the Dalitz plot, which is enriched in \Dz mesons that did not undergo oscillations. We neglect the decay-time resolutions, which are typically $0.1\tau$ and $0.25\tau$ for the \PromptDecay and \SLDecay samples, respectively; and account for this approximation in the systematic uncertainties.

The mixing parameters are determined by minimizing a least-squares function that compares the decay-time evolution of signal yields ($N$) observed in Dalitz bins $-b$ and $+b$, along with their uncertainties ($\sigma$), with the expected values reported in \cref{eq:bin-flip-ratio},
\begin{align}
\chisq \equiv& \sum_{\rm pr,\,sl}\sum_{\rm l,\,d}\sum_{+, -}\sum_{b,j}\frac{(N^\pm_{-bj}-N^\pm_{+bj}R^\pm_{+bj})^2}{(\sigma^\pm_{-bj})^2+(\sigma^\pm_{+bj}R_{+bj}^\pm)^2}\nonumber\\
&+\sum_{b,b'}\left(X^{\rm{CLEO}}_b-X_b\right)(V_{\rm{CLEO}}^{-1})_{bb'}\left(X^{\rm{CLEO}}_{b'}-X_{b'}\right).\label{eq:fit-chi2}
\end{align}
We fit simultaneously the prompt (pr) and semileptonic (sl) samples, separated between long (l) and downstream (d) \KS candidates, as well as between \Dz ($+$) and \Dzb ($-$) flavors, across all decay-time bins $j$ and Dalitz-plot bins $b$. We constrain the parameters $X_b$ to the values $X_b^{\rm{CLEO}}$ measured by the CLEO collaboration through a Gaussian penalty term that uses the sum $V_{\rm{CLEO}}$ of the statistical and systematic covariance matrices~\cite{Libby:2010nu}. In the fit, the parameters $r_b$ are determined independently for each subsample (pr, sl, l, d) because they are affected by the sample-specific variation of the efficiency over the Dalitz plot~\cite{binflip-paper}. The values of \xcp, \deltax, and \deltay were kept blind until the analysis was finalized.

\begin{table}[t]
\centering
\caption{Fit results. The first contribution to the uncertainty is statistical, and the second is systematic.}\label{tab:fit-results}
\begin{tabular}{lcrrrrrr}
\toprule
Parameter & Value & \multicolumn{3}{c}{Stat.\ correlations} & \multicolumn{3}{c}{Syst.\ correlations}\\
 & [$10^{-3}$] & \ycp & \deltax & \deltay & \ycp & \deltax & \deltay\\
\midrule
\xcp & $\phantom{-}\xcpRes\phantom{0}\pm\xcpStat\phantom{0}\pm\xcpSyst\phantom{0}$  & $-0.17$ &  $0.04$ & $-0.02$ &  $0.15$ &  $0.01$ & $-0.02$\\
\ycp & $\phantom{-}\ycpRes\phantom{0}\pm\ycpStat\phantom{0}\pm\ycpSyst\phantom{0}$  &         & $-0.03$ &  $0.01$ &         & $-0.05$ & $-0.03$\\
\deltax & $\dxRes\pm\dxStat\pm\dxSyst$  &         &         & $-0.13$ &         &         &  $0.14$\\
\deltay & $\phantom{-}\dyRes\phantom{0}\pm\dyStat\phantom{0}\pm\dySyst\phantom{0}$  &         &         & \\
\bottomrule
\end{tabular}
\end{table}

\begin{figure}[ht]
\centering
\includegraphics[width=0.6\textwidth]{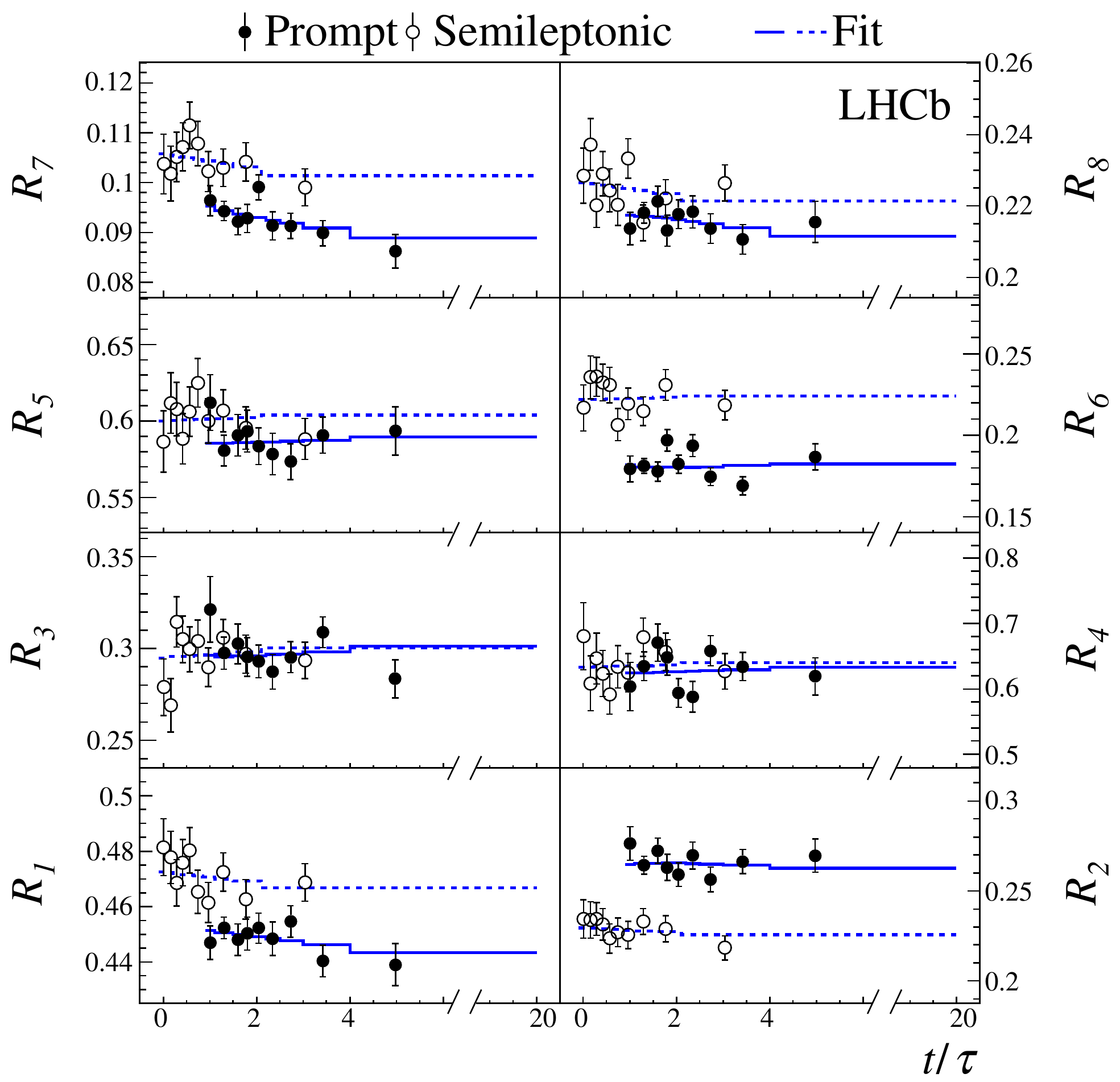}\\
\includegraphics[width=0.6\textwidth]{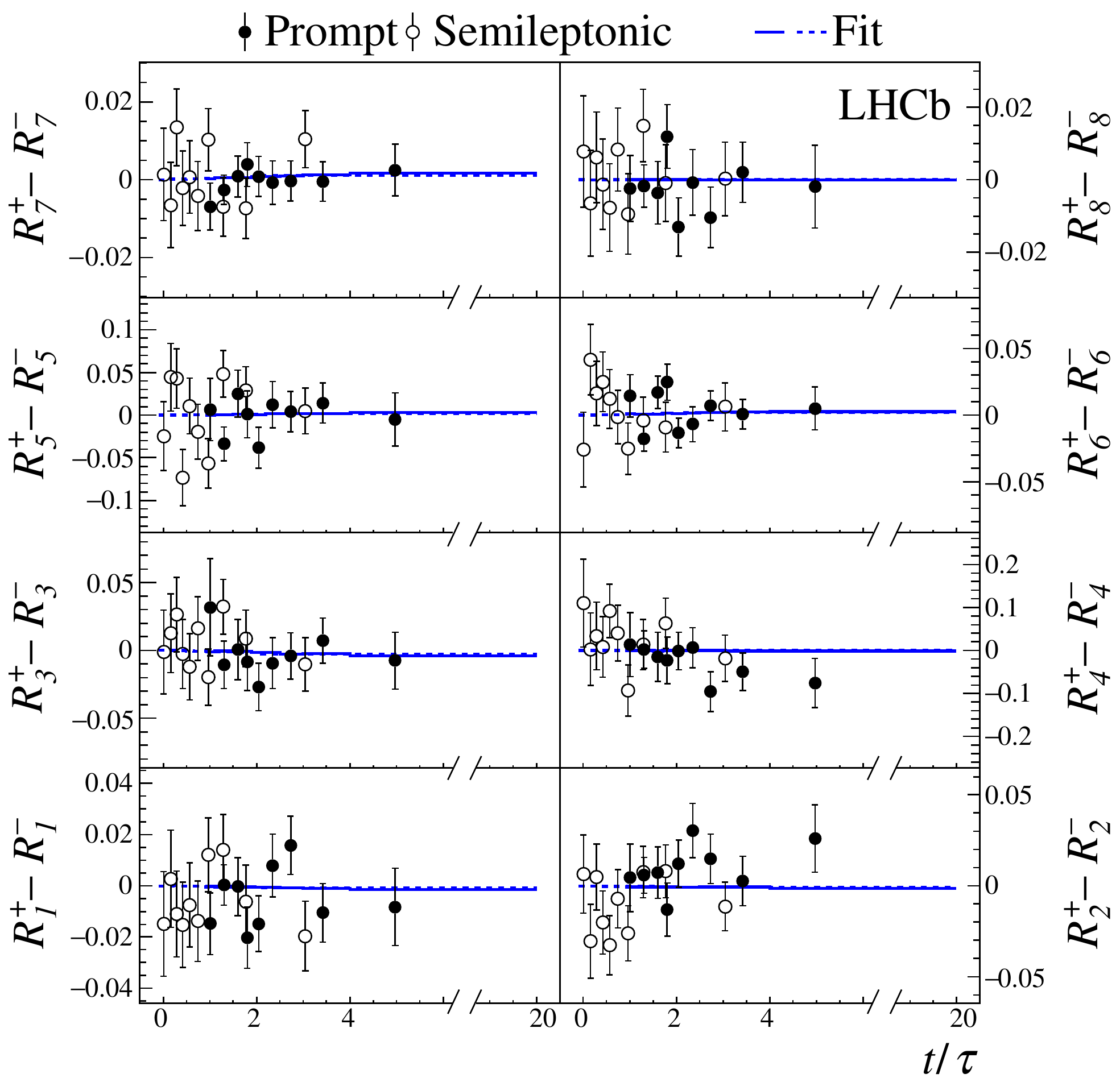}\\
\caption{(Top) \CP-averaged yield ratios and (bottom) differences of \Dz and \Dzb yield ratios as functions of $t/\tau$ for each Dalitz bin. Prompt (closed points) and semileptonic (open points) data are shown separately. Fit projections over the prompt (solid line) and semileptonic (dashed line) data are overlaid.\label{fig:fit-results}}
\end{figure}

\cref{fig:fit-results} shows the yield ratios with fit projections overlaid for prompt and semileptonic data. The offsets between semileptonic and prompt data are due to sample-specific efficiency variations across the Dalitz plot; their slopes, due to charm oscillations, are consistent across samples. \cref{tab:fit-results} lists the results. The data are consistent with \CP symmetry ($\deltax=\deltay=0$). The precision is dominated by the statistical contribution, which incorporates a subleading component due to the precision of the CLEO measurements.

The dominant systematic uncertainties on \xcp are associated with the 3\% contamination from secondary \Dstarp decays in the prompt sample ($0.24\times10^{-3}$) and from the 1\% contamination of genuine \Dz mesons associated with random muons in the semileptonic sample ($0.34\times10^{-3}$). Biases due to the neglected decay-time and $m_\pm^2$ resolutions, and the neglected efficiency variations across decay time and Dalitz plot, constitute the dominant systematic uncertainty on \ycp ($0.94\times10^{-3}$). Possible asymmetric nonuniformities with respect to the bisector in the Dalitz plot induced by reconstruction inefficiencies dominate the systematic uncertainty on \deltax ($0.22\times10^{-3}$) and \deltay ($0.25\times10^{-3}$). Other minor effects, such as mismodeling in the signal-yield fits or in the determination of the bin-averaged decay times, are also considered. The consistency between results on the prompt and semileptonic sample (see \nameref{supplemental-material}), and on various partitions of the data, supports the robustness of the analysis,  including the correction of the $\left(\mpps,t\right)$ correlations.

In summary, we report a measurement of the normalized mass difference between neutral charm-meson eigenstates using the recently proposed bin-flip method. Allowing for \CP violation in charm mixing, or in the interference between mixing and decay, we measure the \CP-averaged normalized mass difference \mbox{$\xcp = [\xcpRes\pm\xcpStat\stat\pm\xcpSyst\syst]\xUnits$}, and the \CP-violating parameter \mbox{$\deltax = [\dxRes\pm\dxStat\stat\pm\dxSyst\syst]\dxUnits$}. In addition, we report the \CP-averaged normalized width difference \mbox{$\ycp = [\ycpRes\pm\ycpStat\stat\pm\ycpSyst\syst]\yUnits$}, along with the corresponding \CP-violating parameter \mbox{$\deltay = [\dyRes\pm\dyStat\stat\pm\dySyst\syst]\dyUnits$}. We use the results to form a likelihood function of $x$, $y$, $|q/p|$, and $\phi$; and we derive confidence intervals (\cref{tab:derived-results}) using a likelihood-ratio ordering that assumes the observed correlations to be independent of the true parameter values~\cite{LHCb-PAPER-2013-020}. The resulting determination of the mass difference is the most precise from a single experiment, as are the determinations of the \CP-violation parameters. Although our result is consistent with $x=0$ within two standard deviations, combined with the current global knowledge, it yields $x = (3.9\,^{+\,1.1}_{-\,1.2})\times10^{-3}$~\cite{HFLAV16}, strongly contributing to the emerging evidence for a nonzero (positive) mass difference between the neutral charm-meson eigenstates. The global constraints on \CP violation in the \Dz-\Dzb system are also greatly improved, with precisions on $|q/p|$ and $\phi$ more than doubled as compared to previous averages~\cite{HFLAV16}.

\begin{table}[t]
\centering
\caption{Point estimates and 95.5\% confidence-level (CL) intervals for the derived parameters. Uncertainties include statistical and systematic contributions.\label{tab:derived-results}}
\begin{tabular}{lcc}
\toprule
Parameter & Value & 95.5\% CL interval \\
\midrule
$x$ [$10^{-2}$] & $\phantom{-}0.27\,_{-\,0.15}^{+\,0.17}$ & $[-0.05,0.60]$ \\
$y$ [$10^{-2}$] & $\quad\,\,0.74\!\pm\!0.37$ & $[\phantom{-}0.00,1.50]$ \\
$|q/p|$         & $\phantom{-}1.05\,_{-\,0.17}^{+\,0.22}$ & $[\phantom{-}0.55,2.15]$ \\
$\phi$          & $-0.09\,_{-\,0.16}^{+\,0.11}$ & $[-0.73,0.29]$ \\
\bottomrule
\end{tabular}
\end{table}

%% file: acknowledgements.tex
\section*{Acknowledgements}
%
%
\noindent We express our gratitude to our colleagues in the CERN
accelerator departments for the excellent performance of the LHC. We
thank the technical and administrative staff at the LHCb
institutes.
We acknowledge support from CERN and from the national agencies:
CAPES, CNPq, FAPERJ and FINEP (Brazil); 
MOST and NSFC (China); 
CNRS/IN2P3 (France); 
BMBF, DFG and MPG (Germany); 
INFN (Italy); 
NWO (Netherlands); 
MNiSW and NCN (Poland); 
MEN/IFA (Romania); 
MSHE (Russia); 
MinECo (Spain); 
SNSF and SER (Switzerland); 
NASU (Ukraine); 
STFC (United Kingdom); 
NSF (USA).
We acknowledge the computing resources that are provided by CERN, IN2P3
(France), KIT and DESY (Germany), INFN (Italy), SURF (Netherlands),
PIC (Spain), GridPP (United Kingdom), RRCKI and Yandex
LLC (Russia), CSCS (Switzerland), IFIN-HH (Romania), CBPF (Brazil),
PL-GRID (Poland) and OSC (USA).
We are indebted to the communities behind the multiple open-source
software packages on which we depend.
Individual groups or members have received support from
AvH Foundation (Germany);
EPLANET, Marie Sk\l{}odowska-Curie Actions and ERC (European Union);
ANR, Labex P2IO and OCEVU, and R\'{e}gion Auvergne-Rh\^{o}ne-Alpes (France);
Key Research Program of Frontier Sciences of CAS, CAS PIFI, and the Thousand Talents Program (China);
RFBR, RSF and Yandex LLC (Russia);
GVA, XuntaGal and GENCAT (Spain);
the Royal Society
and the Leverhulme Trust (United Kingdom);
Laboratory Directed Research and Development program of LANL (USA).

%% file: supplemental-material.tex
\section*{Supplemental material\label{supplemental-material}}

\subsection*{Dalitz-plot distribution and binning scheme}
\begin{figure}[h!!]
\centering
\includegraphics[width=0.6\textwidth]{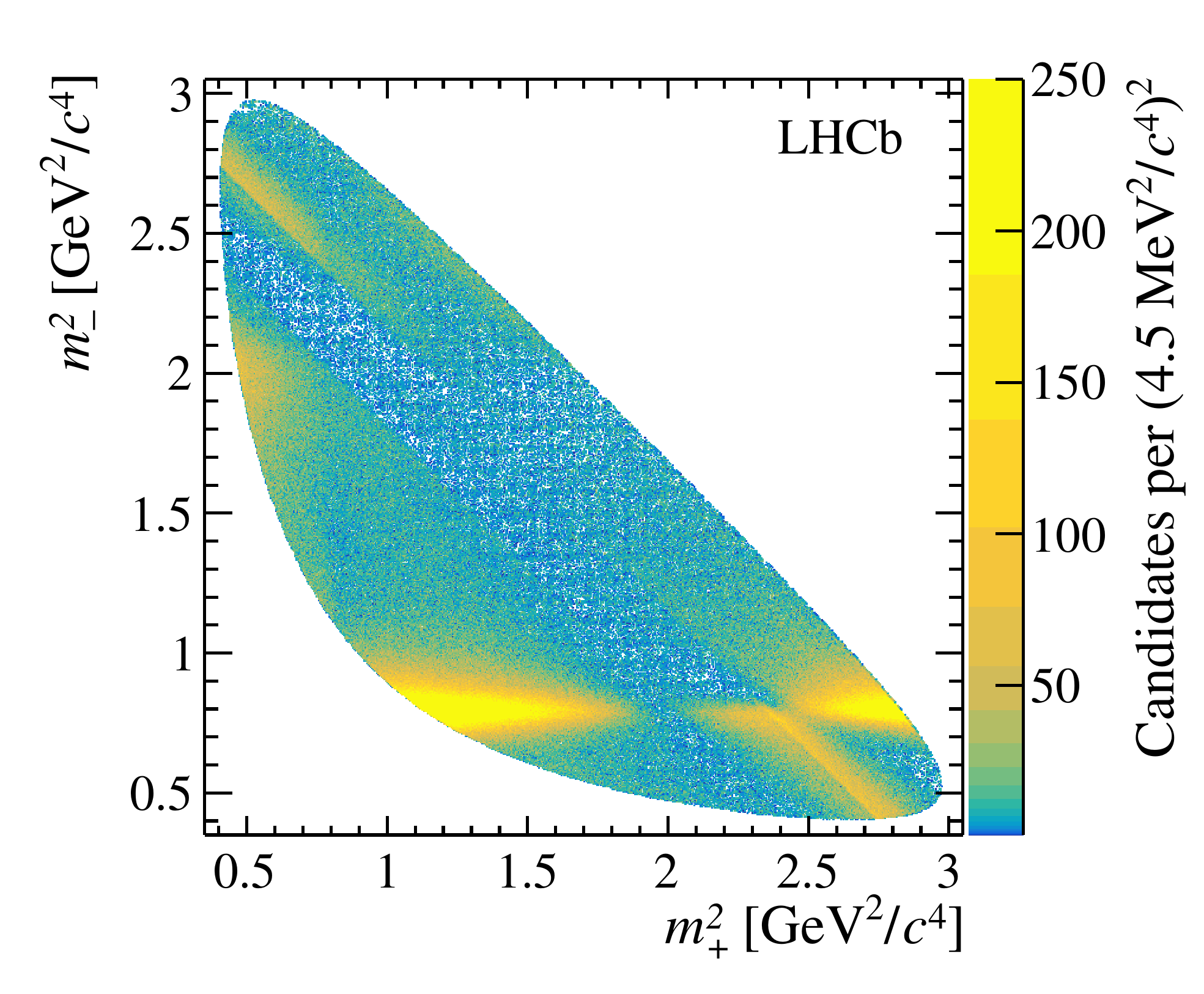}\\
\includegraphics[width=0.6\textwidth]{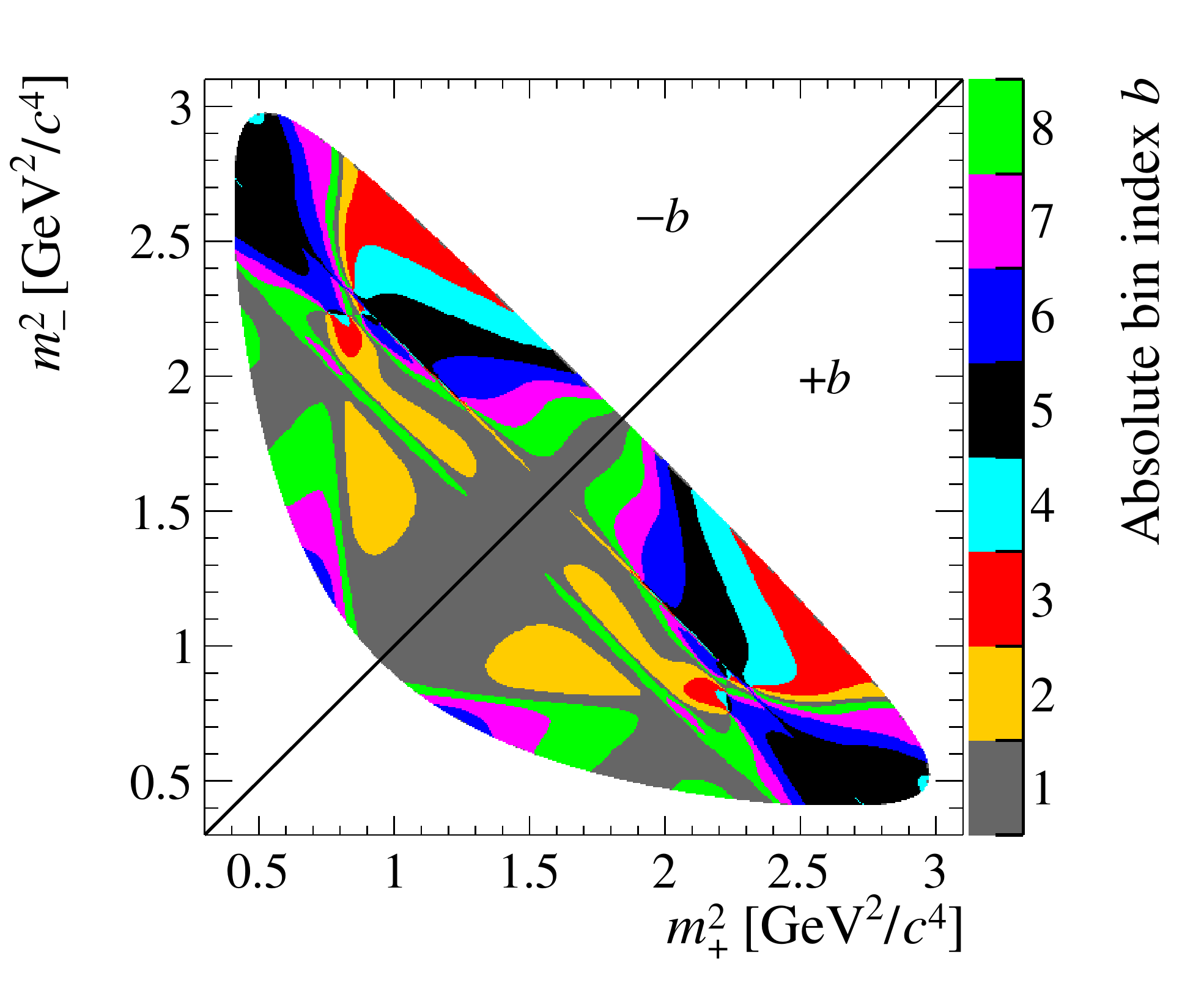}\\
\caption{(Top) Dalitz-plot distribution of background-subtracted \Dkspp candidates. (Bottom) Iso-$\Delta\delta$ binning of the \Dkspp Dalitz plot, reproduced from \href{https://doi.org/10.1103/PhysRevD.82.112006}{Phys.\ Rev.\ {\bf D82} (2010) 112006}. The bins are symmetric with respect to the $m_+^2=m_-^2$ bisector; positive indices refer to bins in the (lower) $m_+^2 > m_-^2$ region; negative indices refer to those in the (upper) $m_+^2 < m_-^2$ region. Colors indicate the absolute value of the bin index $b$.\label{fig:dalitz}}
\end{figure}

\cref{fig:dalitz} shows the decay-time-integrated Dalitz-plot distribution of the background-subtracted \Dkspp candidates used in the analysis, together with the Dalitz-plot binning scheme. No efficiency corrections are applied. All samples are combined.

\subsection*{Prompt- and semileptonic-only fit results}
\cref{tab:prompt-only,tab:sl-only} report results obtained by fitting independently the prompt and semileptonic data samples, respectively.

\begin{table}[h!!]
\centering
\caption{Results of the fit to the prompt data sample. The first contribution to the uncertainty is statistical, the second systematic.\label{tab:prompt-only}}
\begin{tabular}{lcrrrrrr}
\toprule
Parameter & Value & \multicolumn{3}{c}{Stat.\ correlations} & \multicolumn{3}{c}{Syst.\ correlations}\\
 & [$10^{-3}$] & \ycp & \deltax & \deltay & \ycp & \deltax & \deltay\\
\midrule
\xcp & $\phantom{-}3.0\phantom{0}\pm1.9\phantom{0}\pm0.5\phantom{0}$  & $-0.14$ &  $0.05$ & $-0.04$ &  $0.22$ &  $0.00$ & $-0.01$\\
\ycp & $\phantom{-}6.5\phantom{0}\pm4.3\phantom{0}\pm1.5\phantom{0}$  &         & $-0.03$ &  $0.02$ &         & $-0.05$ & $-0.05$\\
\deltax & $-0.41\pm0.78\pm0.24$  &         &         & $-0.12$ &         &         &  $0.17$\\
\deltay & $\phantom{-}0.3\phantom{0}\pm1.8\phantom{0}\pm0.3\phantom{0}$  &         &         & \\
\bottomrule
\end{tabular}
\end{table}

\begin{table}[h!!]
\centering
\caption{Results of the fit to the semileptonic data sample. The first contribution to the uncertainty is statistical, the second systematic.\label{tab:sl-only}}
\begin{tabular}{lcrrrrrr}
\toprule
Parameter & Value & \multicolumn{3}{c}{Stat.\ correlations} & \multicolumn{3}{c}{Syst.\ correlations}\\
 & [$10^{-3}$] & \ycp & \deltax & \deltay & \ycp & \deltax & \deltay\\
\midrule
\xcp    & $\phantom{-}2.3\pm2.8\pm1.0$ & $-0.08$ & $0.03$ &  $0.03$ &  $0.14$ &  $0.00$ & $-0.03$\\
\ycp    & $\phantom{-}8.5\pm6.3\pm0.7$ &         & $0.01$ &  $0.02$ &         & $-0.06$ & $-0.03$\\
\deltax &           $-0.9\pm1.7\pm0.3$ &         &        & $-0.07$ &         &         &  $0.13$\\
\deltay & $\phantom{-}2.2\pm4.0\pm0.2$ &         &        & \\
\bottomrule
\end{tabular}
\end{table}

%% file: LHCb_AuthorsList/LHCb_Authorship_15-Jan-2019.tex
\centerline
{\large\bf LHCb Collaboration}
\begin
{flushleft}
\small
R.~Aaij$^{29}$,
C.~Abell{\'a}n~Beteta$^{46}$,
B.~Adeva$^{43}$,
M.~Adinolfi$^{50}$,
C.A.~Aidala$^{77}$,
Z.~Ajaltouni$^{7}$,
S.~Akar$^{61}$,
P.~Albicocco$^{20}$,
J.~Albrecht$^{12}$,
F.~Alessio$^{44}$
% ...

{\footnotesize \it

% ...
$ ^{7}$Universit{\'e} Clermont Auvergne, CNRS/IN2P3, LPC, Clermont-Ferrand, France\\
% ...
$ ^{12}$Fakult{\"a}t Physik, Technische Universit{\"a}t Dortmund, Dortmund, Germany\\
% ...
$ ^{20}$INFN Laboratori Nazionali di Frascati, Frascati, Italy\\
% ...
$ ^{29}$Nikhef National Institute for Subatomic Physics, Amsterdam, Netherlands\\
% ...
$ ^{43}$Instituto Galego de F{\'\i}sica de Altas Enerx{\'\i}as (IGFAE), Universidade de Santiago de Compostela, Santiago de Compostela, Spain\\
$ ^{44}$European Organization for Nuclear Research (CERN), Geneva, Switzerland\\
% ...
$ ^{46}$Physik-Institut, Universit{\"a}t Z{\"u}rich, Z{\"u}rich, Switzerland\\
% ...
$ ^{50}$H.H. Wills Physics Laboratory, University of Bristol, Bristol, United Kingdom\\
% ...
$ ^{61}$University of Cincinnati, Cincinnati, OH, United States\\
% ...
$ ^{63}$Syracuse University, Syracuse, NY, United States\\
% ...
$ ^{77}$University of Michigan, Ann Arbor, United States, associated to $^{63}$\\
% ...
}
\end{flushleft}